\documentclass{article}

\PassOptionsToPackage{numbers,sort}{natbib}
\usepackage{listings}
\usepackage{xcolor}
\usepackage{float}
\usepackage{booktabs}
\usepackage{tabularx}
\usepackage{longtable} %
\usepackage{array}     %
\usepackage{placeins}  %
\usepackage{amsmath}
\usepackage{bm}
\usepackage{chngcntr}
\lstdefinestyle{agenttrace}{
  basicstyle=\ttfamily\footnotesize,
  breaklines=true,
  breakatwhitespace=false,
  columns=fullflexible,
  keepspaces=true,
  frame=single,
  backgroundcolor=\color{gray!8},
  rulecolor=\color{gray!50},
  framesep=6pt,
  xleftmargin=4pt,
  xrightmargin=4pt,
  showstringspaces=false
}
\usepackage[final]{neurips_2026}
\makeatletter
\renewcommand{\@noticestring}{Workshop on Who Verifies the Agents? Toward Reliable Agent Development, NeurIPS 2026.}
\makeatother

\usepackage[utf8]{inputenc} %
\usepackage[T1]{fontenc}    %
\usepackage{hyperref}       %
\usepackage{url}            %
\usepackage{xurl}           %
\usepackage{booktabs}       %
\usepackage{amsfonts}       %
\usepackage{amsmath}
\usepackage{nicefrac}       %
\usepackage{microtype}      %
\usepackage{xcolor}         %
\usepackage{graphicx}       %
\usepackage{xfrac}          %

\title{Can Coding Agents Write and Transpile Unit Tests?}

\author{%
Gil Pasternak\thanks{Equal contribution.} \\
Code Metal \\
\texttt{gil.pasternak@codemetal.ai} \\
\And
AJ Langley\footnotemark[1] \\
Code Metal \\
\texttt{aj.langley@codemetal.ai} \\
\And
Tom Yotam\footnotemark[1] \\
Code Metal \\
\texttt{tom@codemetal.ai} \\
\And
Ellie Kitanidis\\
Code Metal \\
\texttt{ellie@codemetal.ai} \\
}

\begin{document}

\begingroup
\renewcommand{\thefootnote}{\fnsymbol{footnote}}
\maketitle
\endgroup
\setcounter{footnote}{0}
\begin{abstract} Coding assistants built on Large Language Models (LLMs) have become a staple of modern software development, increasingly trusted to execute workflows and modernize code end-to-end. Much of this trust rests on passing tests, which are often written by the coding agents themselves. In this work, we examine the ability of these agents to both write reliable tests and translate them between programming languages. We benchmark three popular coding agents, each evaluated with a native frontier model and high reasoning effort. Verifying across a broad suite of metrics, we find that coding agents far exceed human written suites on traditional measures of coverage. This advantage, however, is uneven: agent test quality degrades sharply with project size, and much of the advantage lies in a greater number of written tests. We also show that agentic test translation---while seemingly strong in aggregate---remains inconsistent and unreliable.
\end{abstract}

\section{Introduction}
Coding agents have become integral to modern software development, increasingly trusted with tasks as ambitious as generating and transpiling (translating) entire codebases. Verifying this work, however, requires well-constructed tests, which are often missing in realistic code generation settings. Translation is similarly afflicted: even when source language suites are reliable, producing an equally reliable suite in the target language is difficult, as each language introduces distinct translation edge cases and failure modes.

Prior work shows that LLMs can generally write unit tests for small, isolated tasks \cite{hassan2026python, schafer2024empirical}. By contrast, the quality of generated tests at the scale of full codebases, across different programming languages, and using state-of-the-art coding assistants has received considerably less attention. Moreover, even as agents are increasingly tasked with porting legacy systems to newer languages \cite{moti2026legacytranslate, gaudl2026reengineering, bhatnagar2026modernization}, their ability to faithfully translate tests remains underexplored.

\begin{figure*}[h]
  \centering
  \includegraphics[width=0.9\textwidth,trim=0 0.5cm 0 0.5cm,clip]{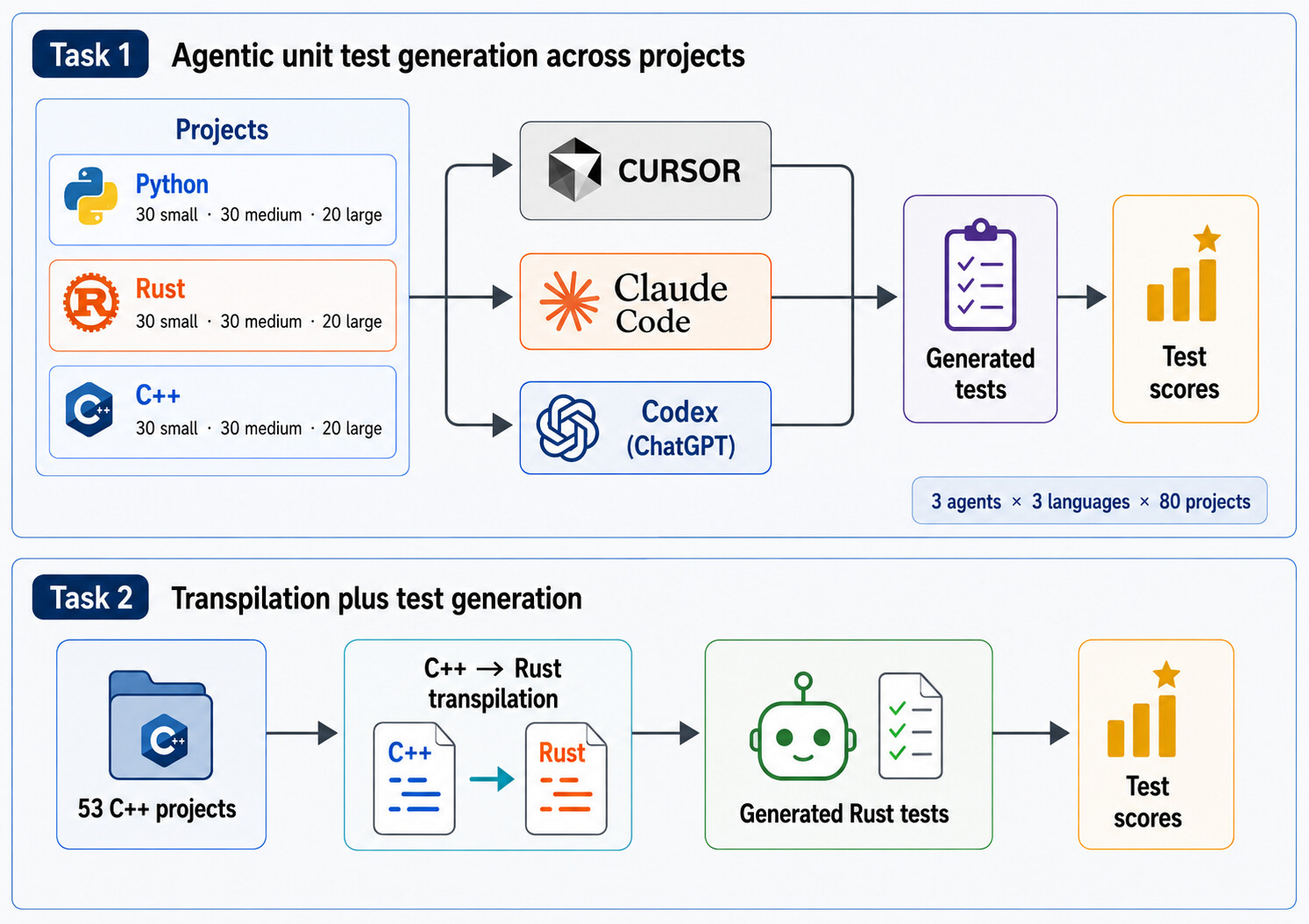}
  \caption{\textbf{Experimental Design Overview}. \textbf{(Top):} We use Cursor, Claude Code, and Codex to generate unit tests for 240 projects across three languages, measuring unit testing capabilities. \textbf{(Bottom)} Additionally, we assess test translation from C++ to Rust, evaluating the model's ability to translate existing C++ test suites.}
  \label{fig:design}
\end{figure*}

In this work, we assemble a dataset of 240 open source projects with human written tests across C++, Python, and Rust. We then use this dataset to evaluate the test generation and translation abilities of three popular coding frameworks: Claude Code, Codex, and Cursor. Combining a broad suite of metrics, agent trace analyses, and controlled experiments, we characterize the strengths and weaknesses of these frameworks. Our contributions are as follows:

\begin{enumerate}
    \item We conduct a detailed investigation of the capabilities and failure modes of modern agentic test generation, finding that coding agents write far more comprehensive suites than their measured human counterparts. However, they often write less thorough individual tests.
    \item We extend our analysis to test transpilation, showing that agents are inconsistent and unfaithful test transpilers in a way that aggregate metrics lack the power to detect. 
    \item We perform an analysis of agentic reasoning traces, exposing pitfalls that sabotage unit test generation and revealing avenues for future work.
\end{enumerate}

\section{Data}
Two datasets underpin our study: a corpus of 240 permissively licensed GitHub projects for the test generation task, and a 50-project subset enhanced for transpilation. Both datasets are described below.

\subsection{Unit testing data}
We collect 240 projects, stratified uniformly across Python, C++, and Rust. Within each language, we further stratify by size: 30 small projects (0-1,000 lines of source code), 30 medium (1k-5k), and 20 large (5k-10k). All collected projects meet the following criteria:

\begin{itemize}
    \item Permissive licensing for redistribution (Apache 2.0 or MIT)
    \item A fully passing and present test suite
    \item C++ datasets: a CMake build (\texttt{CMakeLists.txt}) and no external dependencies
\end{itemize}

We place additional restrictions on C++ projects since they will be translated in our test transpilation experiment, and many dependencies do not have an obvious Rust counterpart.

Our preprocessing pipeline refactors all projects into a uniform format. It then anonymizes them by applying semantic preserving changes to naming schemes and deleting non-docstring documentation. This serves to differentiate the data from that which may have been seen at training time, thus reducing replication of memorized human tests \cite{haroon2026evolution}. The dataset statistics are provided in Table~\ref{tab:argus-corpus}.

\begin{figure*}[t]
  \centering
  \includegraphics[width=0.95\textwidth,height=0.55\textheight,keepaspectratio]{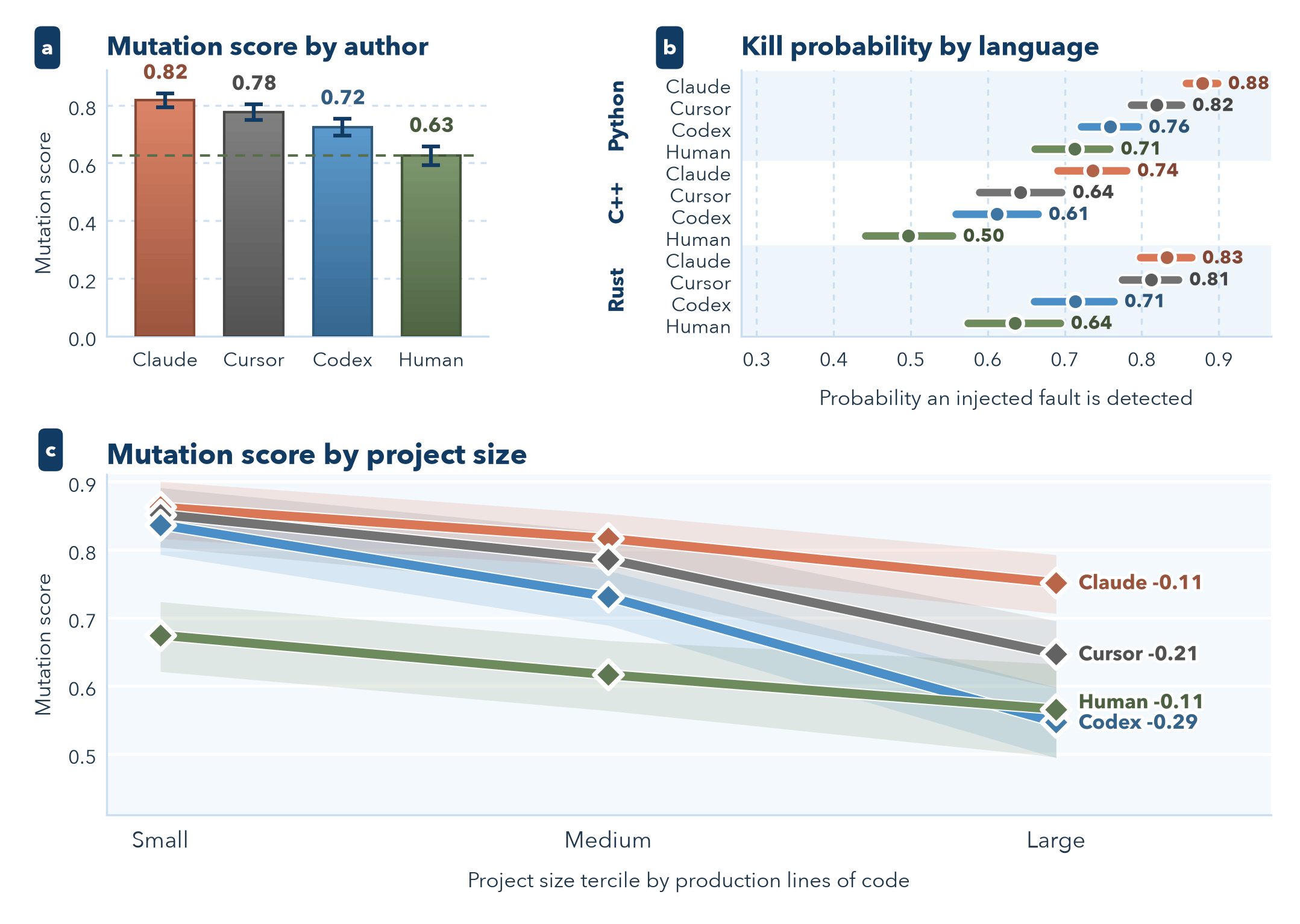}
\caption{\textbf{Coding agents outperform human-written test suites on
average, but the lead narrows on larger projects.}
\textbf{(a)}~Mutation score by suite author; all agents exceed
human baseline.
\textbf{(b)}~Mutant kill probability by language. Measured suites perform best in Python and worst in C++ across the board.
\textbf{(c)}~Mutation score by project-size. Agent scores decay
faster than human scores; on large projects only Claude is
far ahead. Error bands are 95\% CIs.}
  \label{fig:overview}
\end{figure*}

\begin{table}[t]
\centering
\caption{Unit Test generation corpus: 240 frozen, anonymized repositories,
stratified by language and size.}
\label{tab:argus-corpus}
\begin{tabular}{lrcccrc}
\toprule
\textbf{Language} & $\bm{N}$ & \textbf{Small} & \textbf{Medium} & \textbf{Large} & \textbf{LOC med.\ [IQR]} & \textbf{Files med.} \\
\midrule
Python & 80 & 30 & 30 & 20 & $1{,}656$ [$738$, $5{,}111$] & 8 \\
C++    & 80 & 30 & 30 & 20 & $1{,}408$ [$615$, $5{,}122$] & 5 \\
Rust   & 80 & 30 & 30 & 20 & $2{,}526$ [$734$, $5{,}012$] & 8.5 \\
\midrule
Total  & 240 & 90 & 90 & 60 & $1{,}688$ [$700$, $5{,}015$] & 8 \\
\bottomrule
\end{tabular}
\end{table}

\subsection{Transpilation Data}
\label{sec:data:transpiled}
From the C++ corpus, we transpile 50 projects to Rust and release both the source and transpiled artifacts. These projects map cleanly, having platform-agnostic builds and no dependencies that might lack a Rust equivalent. Our final dataset consists of 23 small projects, 20 medium projects, and seven large projects. 

\section{Metrics and measurement}
\label{sec:metrics}

We score every test suite against a fixed per-project checklist: the production lines that exist, the mutants (small bugs) we inject, and the project's public methods. The first two anchor our traditional software testing metrics (line coverage, mutation score, and vacuous test rate) which capture whether tests execute the code and whether they can detect basic faults, but not whether they validate edge cases and core behaviors. For that, we introduce an LLM-as-judge protocol that scores suites against a catalog of important behaviors and edge cases, enabling finer-grained semantic comparison.

We measure the project's own tests once, then score agent tests against the same reference. When a score would divide by zero, we omit that project from the average.

\paragraph{Line coverage.}
Let $U$ be the set of all measurable production lines in a given project. A line
is considered "covered" only if it is run by a test. We only keep tests that pass,
and we drop tests that crash, fail spuriously, or only pass when run in conjunction with other tests. If the suite never runs, or no test passes, the
score is $0$:
\begin{equation}
  C
  = \frac{\lvert\mathrm{Cover}(P)\rvert}{\lvert U\rvert}.
\end{equation}
$\mathrm{Cover}(P)$ is the lines the remaining passing tests $P$ ran.

\paragraph{Mutation score.}
We make small automatic mutations (bug-inducing edits) to the production code, such as flipping a comparison, swapping $+$ and $-$, negating a condition, or changing a number $c$ to $c{+}1$ or $0$. Each project gets a fixed sample of at most 120 such edits; those that do not compile are thrown out. The tests \emph{kill} a mutation if some test fails, hangs, or crashes when running against it.
\begin{equation}
  \mathrm{MS}
  = \frac{\lvert K\rvert}{\lvert V\rvert}.
\end{equation}
$K$ is the set of mutants the tests kill. $V$ is the set of edits that compiled.

\paragraph{Vacuous-test rate.}
We replace each function body with a do-nothing stub
(\texttt{return None} in Python, an empty, error raising, or default return in C++ and
Rust). We then define the vacuous test rate by the fraction of tests that continue to pass when run against the stubbed implementation (and were thus not validating behavior in the first place):
\begin{equation}
  R_{\mathrm{vac}}
  = \frac{\lvert A\setminus\mathit{Sens}\rvert}{\lvert A\rvert}.
\end{equation}
Where $A$ is the set of tests that pass on the original (non-stubbed)
implementation. $\mathit{Sens}$ is the subset of $A$ that then fail, hang,
or crash when run against the stubbed implementation.
\paragraph{LLM-as-judge.}
Once per project, a judge model (Kimi~K3) reads the public functions
and writes a list of important behaviors and edge case inputs for these behaviors. Every
suite is scored against that same list. A behavior counts as covered
if a simple text pattern for it appears in the tests. Edge cases need
a second model pass: tests often build such inputs in variables,
and thus may be missed by a fixed pattern.
\begin{equation}
  C_{\mathrm{beh}}
  = \frac{\lvert\mathcal{B}_{\mathrm{hit}}\rvert}{\lvert\mathcal{B}\rvert},
  \qquad
  C_{\mathrm{edge}}
  = \frac{\lvert\mathcal{E}_{\mathrm{hit}}\rvert}{\lvert\mathcal{E}\rvert}.
\end{equation}
$\mathcal{B}$ and $\mathcal{E}$ are the lists of behaviors and edge
cases. The $\mathrm{hit}$ sets are the ones the suite covered.

All remaining metric definitions can be found in Appendix~\ref{app:metrics}.

\section{Experiments}
We run three containerized and sandboxed experiments, building each project inside a clean container in accordance with its build instructions. Unless otherwise stated, we run all models with high reasoning and unlimited token budgets for testing. Our prompts---along with other shared settings---are not specialized to any agent and can be found in Appendix~\ref{prompts}. Individual experiments are detailed in the following sections.

\subsection{Unit testing}
\label{sec:experiments:unit}
To measure the ability of modern coding frameworks to write unit tests, we remove all human tests and instruct the three coding agents to repopulate them. A coding assistant may run the tests and observe output -- it is bound only by a lack of network access so that it cannot look up the projects online. For each coding framework, we leverage the most recent frontier LLM: GPT-5.6-sol for Codex, Claude-Opus-5 for Claude Code, and Cursor-Grok-4.6 for Cursor. 
\subsection{Test transpilation}
Beyond generating unit tests from scratch, we examine coding agents' ability to translate existing test suites across languages. We center evaluations on C++ to Rust, a pairing of particular interest in code modernization \cite{darpa_tractor, khatry2025crustbench}. Using the transpiled Rust project corpus detailed in section \ref{sec:data:transpiled}, the agent is presented with a full C++ project, the native C++ tests, and the transpiled Rust source. It is then prompted to transpile the C++ test suite to Rust. 

As in the previous section, we leverage the most recent frontier LLM for each coding agent. We also perform unconditional generation of a Rust suite (providing only transpiled Rust, without the source code and source test suite) as a means of comparison.

\subsection{Interpretive analysis}
The Interpretive experiment repeats the unit testing setup in \ref{sec:experiments:unit} on a 225 project subset. We leverage a Codex harness and GPT 5.6, investigating reasoning traces to identify core issues in Codex's test generation workflow. We establish three fundamental failure modes and track them across projects by analyzing local reasoning traces, tool calls, and additional harness data.

\begin{figure*}[t]
  \centering
\includegraphics[width=\textwidth,height=0.78\textheight,keepaspectratio]{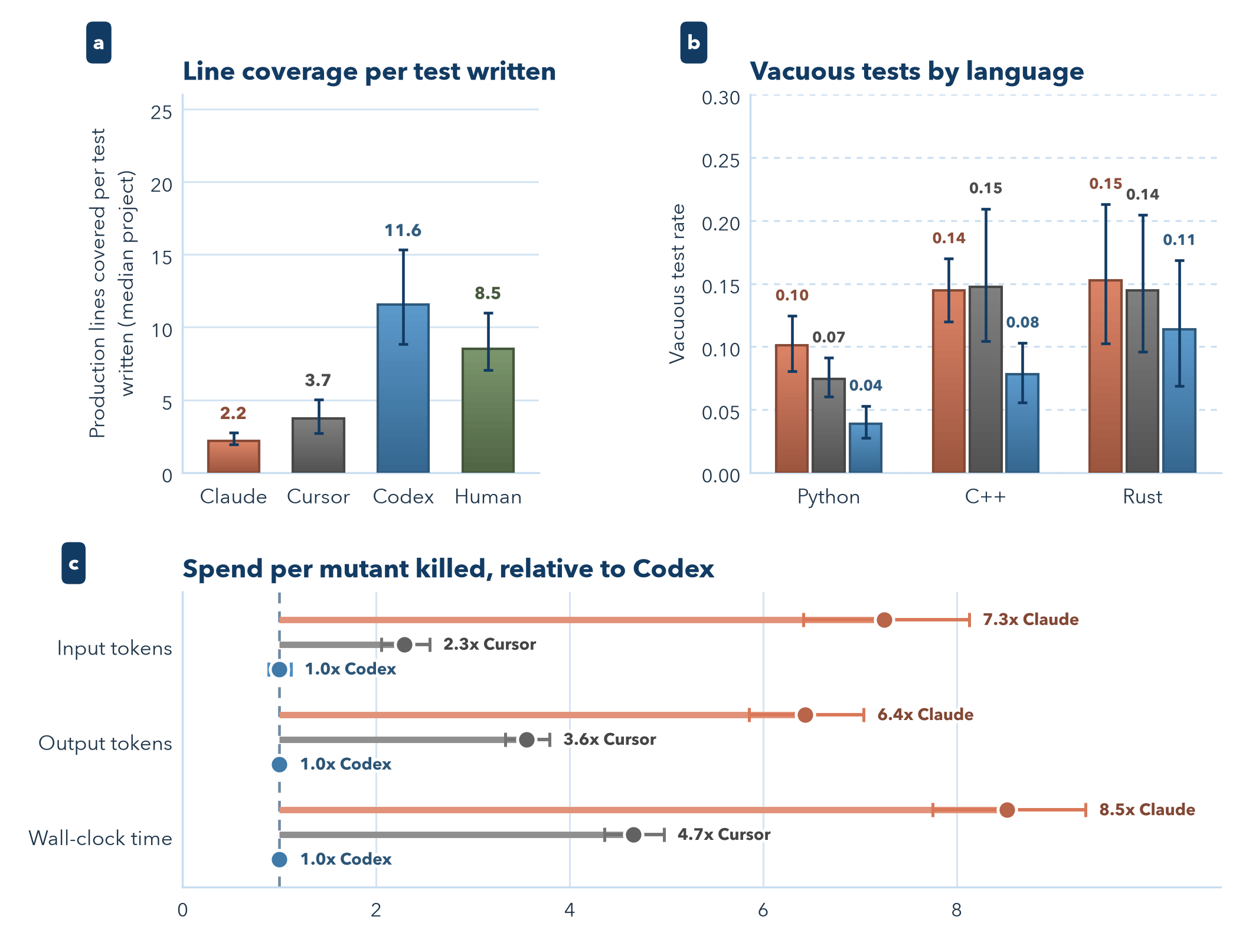}
  \caption{\textbf{Agents achieve similar test quality at very different
costs.}
\textbf{(a)}~Production lines covered per test written: Codex is the
most economical author; Claude reaches comparable coverage with roughly
five times as many tests.
\textbf{(b)}~Vacuous-test rate by language, rising from Python to Rust;
Claude and Cursor fare worst throughout.
\textbf{(c)}~Spend per mutant killed, relative to Codex: Claude uses far more tokens per mutant killed than other suites. Error bars
are 95\% CIs.}
  \label{fig:quality-overview}
\end{figure*}

\section{Results}

\subsection{Unit testing}
Holistically, our results show that the agentic test suites were far more comprehensive than their human counterparts. Figure~\ref{fig:overview} shows statistically significant improvements in mutant kill rate over the human baseline, and the pattern repeats across line coverage, redundancy, behavioral coverage, and edge case coverage.

Stratification, however, exposes clear weaknesses. Agents cover Python capably, but struggle with C++ across the board (Figure~\ref{fig:overview}b). Every agent experiences a dropoff of at least 14\% relative to its Python performance. 

Test quality also decays significantly with project size, and disproportionately so in agents. On large projects, Codex essentially matches the quality of human written suites, in stark contrast to small projects where agents dominate. Full results appear in Appendix ~\ref{tab:all}.

Analyzing suite quality, moreover, conceals a major testing efficiency disparity (Figure~\ref{fig:quality-overview}). On a per-test basis, humans are more efficient authors than two of the three coding agents: each human-written test covers 8.5 production lines, against 3.7 for Cursor and 2.2 for Claude Code, with only Codex ahead at 11.6. (Figure~\ref{fig:quality-overview}a). Claude Code in particular reaches its coverage primarily through volume, typically writing roughly over seven times as many tests as the human baseline. This volume carries some waste. The vacuous-test rate (tests that still pass even once the tested function they cover is stubbed out) climbs steadily from Python through C++ to Rust, and is highest for Claude Code on average (Figure~\ref{fig:quality-overview}b). 

The pattern also extends to resources: per mutant killed, Claude Code spends 7.3$\times$ the input tokens, 6.4$\times$ the output tokens, and 8.5$\times$ the wall-clock time of Codex (Figure~\ref{fig:quality-overview}c). Cursor sits near 2 to 5$\times$ on all three measures. Taken together, the figures suggest that large codebases and lower-resource languages remain harder for agents to unit test; and where agents do hold an advantage, much of it result from simply writing more tests.

\FloatBarrier

\subsection{Test transpilation}
\begin{table}[t]
\centering
\caption{Aggregate metrics for human source test suites and transpiled Rust across all three coding agents. Transpiled test suites lag behind C++ source suites in every major indicator, in particular in line coverage. Values are mean $\pm$ sample standard deviation across three runs.}
\label{tab:transpilation:suite-comparison}
\begin{tabular}{lrrrr}
\toprule
\textbf{Metric} & \textbf{C++ human} & \textbf{Claude Opus 5} & \textbf{Codex gpt-5.6} & \textbf{Cursor grok-4.6} \\
\midrule
Mutation score & 0.559 $\pm$ 0.033 & 0.514 $\pm$ 0.033 & 0.489 $\pm$ 0.035 & 0.513 $\pm$ 0.033 \\
Edge case coverage & 0.340 $\pm$ 0.035 & 0.297 $\pm$ 0.031 & 0.287 $\pm$ 0.032 & 0.289 $\pm$ 0.031 \\
Line coverage & 0.747 $\pm$ 0.034 & 0.651 $\pm$ 0.032 & 0.612 $\pm$ 0.036 & 0.638 $\pm$ 0.034 \\
\bottomrule
\end{tabular}
\end{table}
\begin{figure*}[t]
  \centering
  \includegraphics[width=1.0\textwidth]{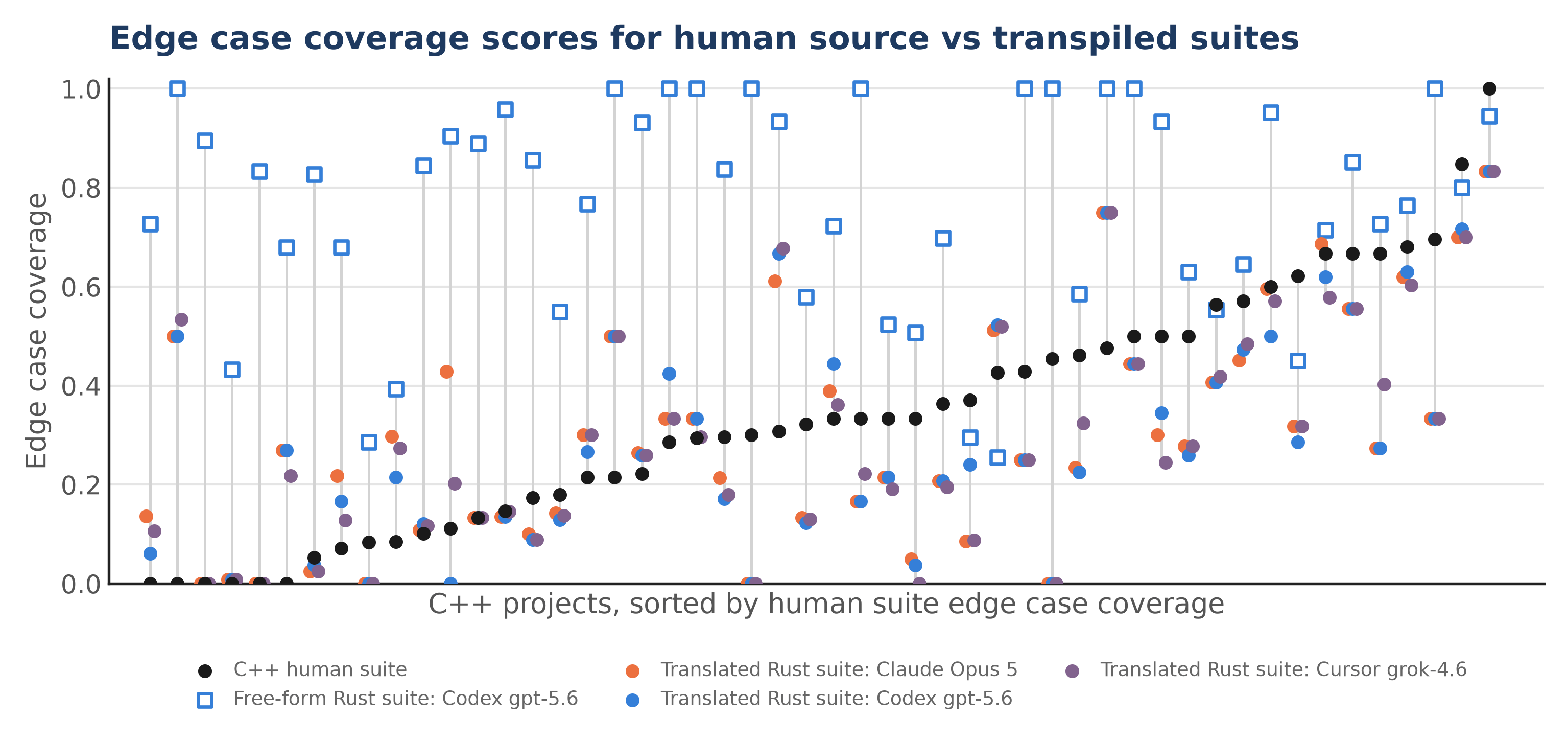}
  \caption{Edge case coverage scores for human source test suites, transpiled Rust test suites, and test suites independently generated from transpiled Rust source.}
  \label{fig:transpilation-edge-case-coverage}
\end{figure*}
\begin{figure*}[t]
  \centering
  \includegraphics[width=1.0\textwidth]{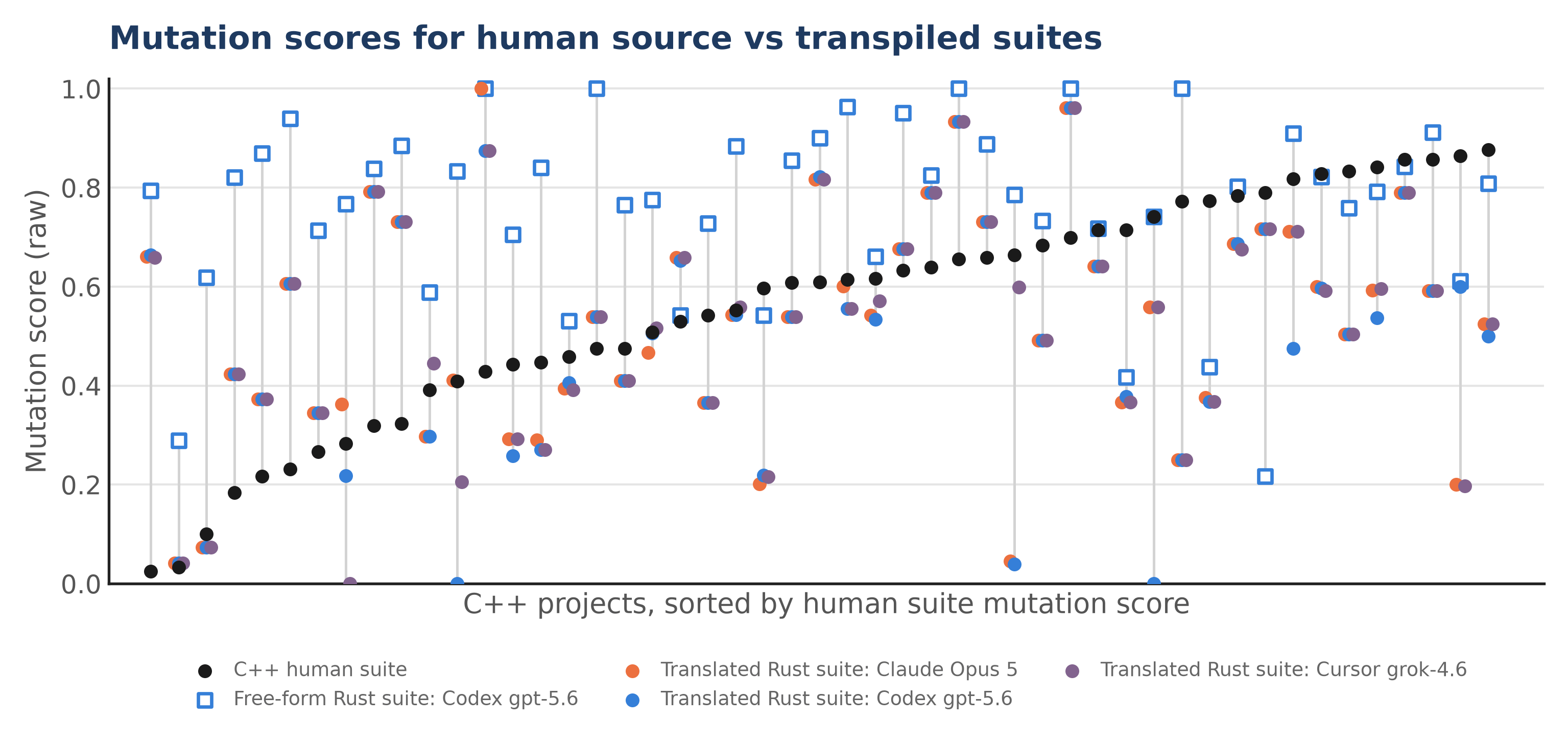}
  \caption{Mutation scores across human, transpiled, and independently generated test suites. Transpilation performance is upper bounded by independently generated performance, and transpiled test suites lag meaningfully behind their human sources on several projects.}
  \label{fig:transpilation-comparison}
\end{figure*}
We evaluate three coding agents---Claude Code, Codex, and Cursor---as test transpilers and place full corpus results in Table~\ref{tab:transpilation:suite-comparison}. The agents appear to be decent transpilers in aggregate, with translated suites roughly maintaining the mutation scores of their C++ baselines. The pointwise picture is different. As seen in Figure~\ref{fig:transpilation-edge-case-coverage}, the agent-transpiled suite underperforms the source suite in edge case coverage 58\% of the time, with dropoffs reaching 45\% in the absolute. In 55\% of projects, the agent kills fewer mutants altogether (Figure~\ref{fig:transpilation-comparison}). In the worst case, the mutant coverage of the source suite experiences nearly a 70\% relative dropoff in all agentic transpilations.

This gap is partially masked by independent test generation. Figure \ref{fig:transpilation-comparison} makes this visible by also plotting suites generated from the final Rust implementation alone, with the C++ source withheld. The comparison reveals a surprising ceiling, showing that free-form generation nearly always upper-bounds the translated suites. In effect, the agent interpolates between translation and generation, sometimes producing tests that far exceed the source suite. However, the coupling cuts both ways. Where unconditional generation is weak, translation is weak and unfaithful. In some instances the translated suite even falls catastrophically below both the free-form suite and the human source. 

Notably, the three agents' translated suites cluster tightly on a significant majority of projects---these failure modes appear to be properties of the task, not of any particular assistant. Taken together, the results raise concerns about the reliability of modern coding agents on transpilation tasks.

\subsection{Interpretive analysis}

We examined 675 test-generation trajectories produced by Codex GPT-5.6 Sol across 225 projects, with three runs per project. The traces reveal failures in repository scope and target selection and in oracle formation, alongside gaps in coverage-feedback utilization. All percentages below use the 675 trajectories as their denominator.

The weakest runs share the following failure modes: skipping relevant code, forming trivial or implementation-anchored oracles, and
failing to implement feedback effectively. These modes appeared even when
the traces showed apparent prior knowledge: the agent identified the upstream library or recalled its structure, but still produced a weak suite
(Listings~\ref{lst:familiarity-cleo}--\ref{lst:familiarity-more-itertools}).

\paragraph{Repository scope and target selection.}
In 103 trajectories (15.3\%), we found evidence that the agent either did not read relevant source code or knowingly omitted a supported testing objective. The traces show relevant implementation files left unread despite their inclusion in the build and their connection to the tested public behavior (Listings~\ref{lst:scope-commands}--\ref{lst:scope-test}). They also contain explicit decisions to exclude identified behavior without retaining an equivalent test.

\paragraph{Oracle formation.}
Seventy-five trajectories (11.1\%) exhibited evidence of a retained oracle-formation failure. These included preserving behavior the agent recognized as erroneous, accepting incidental output from malformed input, and weakening assertions until they no longer established the intended property (Listings~\ref{lst:area-recognition}--\ref{lst:validation-after}). The final suite could therefore pass while retaining assertions that preserved an implementation error or no longer checked the intended validation behavior.

\paragraph{Feedback utilization.}
In 376 trajectories (55.7\%), no coverage-measurement attempt was recorded, either through an explicit agent command or through inherited test infrastructure (Table~\ref{tab:coverage-observations}). The traces show repeated test execution and repair without a corresponding coverage measurement. In these runs, the execution feedback established whether the generated suite passed, but did not measure which implementation paths remained untested (Listings~\ref{lst:feedback-command}--\ref{lst:feedback-result} and Table~\ref{tab:feedback-runs}).

We tested these findings using a failure-aware prompt
(Listing~\ref{lst:failure-aware-prompt}) to run Codex GPT-5.6 Sol on
41 projects spanning multiple languages and sizes. The failure-aware prompt
improves most metrics, at the cost of slightly increased redundancy
(Table~\ref{tab:failure-aware-prompt-comparison}).

\section{Related Works}
\textbf{Datasets and benchmarks.} Existing test generation benchmarks are predominantly in Python or Java. TestEval (210 Python LeetCode problems)~\citep{testeval} and TestBench (108 functions from nine real-world Java projects)~\citep{testbench} provide curated unit-level tasks, but are limited in scale and complexity. TestGenEval~\citep{jain2025testgeneval} and SWT-Bench~\citep{mundler2024swtbench} are substantially larger and more realistic, but are derived from SWE-Bench~\citep{jimenez2024swebench} repository histories. Their target behavior is therefore defined relative to an issue-resolving patch and may span multiple components, making these tasks closer to repository-level regression or issue reproduction than isolated unit testing.

Recent studies also suggest that data contamination and memorized surface patterns can inflate apparent test-generation performance. \citet{huang2025ult} introduce ULT (UnLeakedTestBench), a benchmark of real world Python functions whose corresponding tests do not appear in The Stack v2~\citep{lozhkov2024stackv2}, alongside PLT (PreLeakedTestBench), which contains functions with potentially leaked tests. Across 12 models, average Pass@1 falls from 48.42\% on PLT to 12.69\% on ULT. \citet{haroon2026evolution} reach a related conclusion: even semantics-preserving program transformations substantially reduce generated-test pass rates.

Our dataset instead targets isolated, human written unit tests from 240 real world projects spanning Python, C++, and Rust, with a range of project sizes and complexities. We additionally apply transformations such as identifier renaming to reduce the likelihood that models can rely on memorized public source code or tests.

\textbf{LLMs for test generation.}
\label{sec:related:testgen}
LLMs have been used throughout the unit test generation pipeline. Prior work generates tests from function context with execution feedback~\citep{schafer2024empirical,chen2024chatunitest}, augments existing suites at industrial scale~\citep{alshahwan2024testgenllm}, reproduces bug reports as tests~\citep{kang2023libro}, and improves assertions using retrieved demonstrations~\citep{nashid2023cedar}. Hybrid methods combine LLMs with search-based testing~\citep{lemieux2023codamosa,deljouyi2025utgen} or mutation testing~\citep{dakhel2024mutap}. Others guide generation with program analysis, including method slices~\citep{wang2024hits}, symbolic path constraints~\citep{ryan2024symprompt}, coverage reports~\citep{pizzorno2024coverup}, and uncovered control-flow paths~\citep{gu2026panta}. ASTER extends this style of pipeline across languages~\citep{pan2025aster}.

Empirical studies in JavaScript~\citep{schafer2024empirical}, Java~\citep{siddiq2024junit,yang2024evaluation}, Python~\citep{testeval}, and oracle generation~\citep{molinelli2025oracles} collectively show that LLMs can produce plausible tests for simple methods but degrade substantially on more complex, real world code. Controlled studies identify several recurring failure modes. Generated tests can anchor on an incorrect implementation~\citep{huang2024rethinking}, inherit mistakes from self-written code~\citep{konstantinou2026coding}, or be emptied to make them pass~\citep{bekmyradov2026shortcuts}. Test quality also degrades with program complexity~\citep{abdullin2025testwars} and can remain superficial despite high coverage~\citep{swemutation2026}.

Most of these studies evaluate single model calls or fixed pipelines, often in a single language. Coding agents have received far less attention. SWT-Bench~\citep{mundler2024swtbench} shows that repair agents outperform dedicated test generators at reproducing GitHub issues, but only in Python. The only direct comparison of production assistants we are aware of evaluates 99 LeetCode problems~\citep{hassan2026python}. Whether agents' harnesses, tools, and instructions change these outcomes on real, multilingual repositories remains open.

\textbf{Test migration and cross-language translation.}
\label{sec:related:translation}
LLM code translation is known to be unreliable, with correct translation rates of $\le$ 47\% across models and benchmarks~\citep{pan2024lost}. Test translation has received less attention. Even within a single language, results remain poor: LLMs fail on roughly half of Python \texttt{unittest}-to-\texttt{pytest} migrations~\citep{alves2026migration}. Cross-language migration has so far succeeded only with traditional program analysis and a human in the loop. MUT~\citep{gao2024mut} translates 550 tests through AST rewriting, of which maintainers accepted 253. We are unaware of prior work that measures semantic preservation---including edge cases, assertions, and fault-catching power---when translating tests across languages.
.

\section{Conclusion}
\label{sec:conclusion}
In this paper, we evaluated three coding agents on unit test generation for 240 real-world projects across Python, C++, and Rust. On average, agents beat human suites on traditional metrics, but the gains are uneven: performance degrades with project size and lags on C++. Much of the agents' advantage emanates from writing more tests as opposed to making them stronger individually.

We also study C++$\rightarrow$Rust test transpilation. While aggregate scores look fair, translated suites frequently diverge from their sources and regress on about 55\% of projects. We notice this issue across both the software testing metrics we use and the semantic edge case metrics we define.

Our reasoning-trace analysis suggests three causes for these observations: agents read repositories incompletely, anchor oracles to the implementation, and rarely use coverage feedback. We thus conclude that agent-written, and especially translated, test suites remain unreliable in many scenarios and should be treated as drafts requiring careful review and measurement.

\paragraph{Limitations and future work.} There are several present limitations to our work. First, our corpus consists of open-source repositories and does not include enterprise-scale codebases. Enterprise test suites are often more mature than the baselines we measure here, which could narrow agents' apparent advantage while further stress-testing the reliability of agent-written suites. Building representative enterprise benchmarks, with appropriate privacy and licensing constraints, is an important direction for future work.

Furthermore, we test agentic harnesses with their native models, and do not isolate the individual impact of either the harness or models themselves. This is an important follow-up direction, and could aid practitioners working to implement harness and model level improvements.

Our interpretive analysis is also limited. Reasoning traces come from a single model (GPT) in a single harness (Codex); no comparable traces exist for the other frontier agents in our main experiments. Whether the same failure modes hold across agents, models, harnesses, and project types remains open.

Finally, we document frequent semantic regressions in C++ $\rightarrow$ Rust test transpilation but do not propose a remedy. Future progress likely requires stronger harnesses and evaluation loops, such as verifiers and targeted behavioral checks, alongside post-training tailored to semantic preservation.

\section*{Acknowledgments}
We thank Alecsander Gonzalez for his great help in editing and revising several drafts of this paper.

{\small
\bibliographystyle{plainnat}
\bibliography{related_work}
}

\newpage
\appendix
\section*{Appendix}
\counterwithin{lstlisting}{section}
\lstset{
  basicstyle=\ttfamily\scriptsize,
  breaklines=true,
  breakatwhitespace=false,
  columns=fullflexible,
  keepspaces=true,
  showstringspaces=false,
  frame=single
}
\section{Unit Testing Experiment Metrics}
\label{tab:all}

 \begin{table}[ht]
\centering
\small
\caption{Overall projects: mean of each test-suite author on every metric, and the paired gap between the agents and the human baseline. ``Agent mean'' averages Claude, Cursor, and Codex on the same project.}
\label{tab:metrics-overall}
\begin{tabular}{lrrrrl}
\toprule
 & \multicolumn{4}{c}{Mean by author} & \\
\cmidrule(lr){2-5}
Metric & Claude & Cursor & Codex & Human & Agent mean $-$ Human [95\% CI] \\
\midrule
Mutation score & 0.818 & 0.776 & 0.724 & 0.625 & 0.147 [0.120, 0.174] \\
Line coverage & 0.926 & 0.913 & 0.873 & 0.762 & 0.142 [0.116, 0.168] \\
Coverage at pass & 0.926 & 0.913 & 0.873 & 0.762 & 0.142 [0.116, 0.168] \\
Mutants killed per test & 0.33 & 0.48 & 1.59 & 8.40 & $-$7.63 [$-$10.01, $-$5.25] \\
Assertion density & 0.240 & 0.289 & 0.280 & 0.170 & 0.099 [0.082, 0.116] \\
Test redundancy & 0.160 & 0.175 & 0.150 & 0.253 & $-$0.096 [$-$0.129, $-$0.063] \\
Invalid-test rate & 0.0000 & 0.0017 & 0.0000 & 0.0000 & 0.0006 [$-$0.0002, 0.0013] \\
Stub sensitivity & 0.865 & 0.877 & 0.919 & 0.893 & 0.005 [$-$0.026, 0.036] \\
Vacuous test rate & 0.135 & 0.123 & 0.081 & 0.107 & $-$0.005 [$-$0.036, 0.026] \\
Tests written & 299 & 162 & 47 & 263 & $-$94 [$-$348, 159] \\
Test lines of code & 2277 & 1575 & 536 & 1485 & $-$22 [$-$333, 289] \\
Test lines added & 2277 & 1575 & 536 & -- & -- \\
Assertions written & 543 & 447 & 154 & 217 & 164 [124, 204] \\
Mutants killed per added test & 0.092 & 0.126 & 0.373 & -- & -- \\
Mutants killed per added test line & 0.0101 & 0.0120 & 0.0288 & -- & -- \\
Generation seconds & 1209 & 611 & 121 & -- & -- \\
Tool calls & 68 & 67 & 18 & -- & -- \\
Output tokens & 97774 & 49984 & 12945 & -- & -- \\
Input tokens & 6499378 & 1901918 & 763534 & -- & -- \\
Mutation score per minute & 0.070 & 0.093 & 0.445 & -- & -- \\
Mutation score per tool call & 0.0226 & 0.0156 & 0.0529 & -- & -- \\
Suite runtime seconds & 3.24 & 1.94 & 0.81 & 6.86 & $-$4.87 [$-$8.16, $-$1.57] \\
Behavioral coverage & 0.896 & 0.888 & 0.825 & 0.599 & 0.271 [0.232, 0.310] \\
Edge-case coverage & 0.885 & 0.841 & 0.718 & 0.461 & 0.354 [0.319, 0.388] \\
Behaviors per test & 0.047 & 0.084 & 0.260 & 1.064 & $-$0.937 [$-$1.321, $-$0.552] \\
Edge cases per test & 0.121 & 0.199 & 0.551 & 1.610 & $-$1.323 [$-$1.858, $-$0.787] \\
\bottomrule
\end{tabular}
\end{table}

\newpage

\begin{table}[ht]
\centering
\small
\caption{Python projects: mean of each test-suite author on every metric, and the paired gap between the agents and the human baseline. ``Agent mean'' averages Claude, Cursor, and Codex on the same project.}
\label{tab:metrics-python}
\begin{tabular}{lrrrrl}
\toprule
 & \multicolumn{4}{c}{Mean by author} & \\
\cmidrule(lr){2-5}
Metric & Claude & Cursor & Codex & Human & Agent mean $-$ Human [95\% CI] \\
\midrule
Mutation score & 0.891 & 0.841 & 0.791 & 0.706 & 0.134 [0.082, 0.187] \\
Line coverage & 0.981 & 0.963 & 0.931 & 0.844 & 0.115 [0.072, 0.158] \\
Coverage at pass & 0.981 & 0.963 & 0.931 & 0.844 & 0.115 [0.072, 0.158] \\
Mutants killed per test & 0.23 & 0.39 & 1.11 & 1.41 & $-$0.84 [$-$1.42, $-$0.25] \\
Assertion density & 0.261 & 0.318 & 0.279 & 0.199 & 0.087 [0.060, 0.115] \\
Test redundancy & 0.160 & 0.175 & 0.150 & 0.253 & $-$0.096 [$-$0.129, $-$0.063] \\
Invalid-test rate & 0.0000 & 0.0000 & 0.0000 & 0.0000 & 0.0000 [$-$0.0000, 0.0000] \\
Stub sensitivity & 0.899 & 0.926 & 0.961 & 0.921 & 0.019 [$-$0.039, 0.077] \\
Vacuous test rate & 0.101 & 0.074 & 0.039 & 0.079 & $-$0.019 [$-$0.077, 0.039] \\
Tests written & 514 & 244 & 76 & 679 & $-$401 [$-$1163, 361] \\
Test lines of code & 2262 & 1640 & 599 & 2151 & $-$650 [$-$1335, 35] \\
Test lines added & 2262 & 1640 & 599 & -- & -- \\
Assertions written & 589 & 517 & 172 & 371 & 55 [$-$25, 134] \\
Mutants killed per added test & 0.052 & 0.087 & 0.227 & -- & -- \\
Mutants killed per added test line & 0.0114 & 0.0125 & 0.0296 & -- & -- \\
Generation seconds & 1202 & 663 & 133 & -- & -- \\
Tool calls & 87 & 73 & 20 & -- & -- \\
Output tokens & 96405 & 55670 & 14414 & -- & -- \\
Input tokens & 8483035 & 2277637 & 920344 & -- & -- \\
Mutation score per minute & 0.078 & 0.090 & 0.436 & -- & -- \\
Mutation score per tool call & 0.0194 & 0.0152 & 0.0515 & -- & -- \\
Suite runtime seconds & 3.85 & 3.05 & 1.25 & 9.70 & $-$6.98 [$-$13.02, $-$0.95] \\
Behavioral coverage & 0.915 & 0.909 & 0.858 & 0.623 & 0.271 [0.212, 0.330] \\
Edge-case coverage & 0.932 & 0.886 & 0.764 & 0.525 & 0.336 [0.273, 0.399] \\
Behaviors per test & 0.034 & 0.073 & 0.210 & 0.210 & $-$0.104 [$-$0.240, 0.032] \\
Edge cases per test & 0.096 & 0.183 & 0.462 & 0.281 & $-$0.034 [$-$0.119, 0.051] \\
\bottomrule
\end{tabular}
\end{table}

\newpage

\begin{table}[ht]
\centering
\small
\caption{C++ projects: mean of each test-suite author on every metric, and the paired gap between the agents and the human baseline. ``Agent mean'' averages Claude, Cursor, and Codex on the same project.}
\label{tab:metrics-cpp}
\begin{tabular}{lrrrrl}
\toprule
 & \multicolumn{4}{c}{Mean by author} & \\
\cmidrule(lr){2-5}
Metric & Claude & Cursor & Codex & Human & Agent mean $-$ Human [95\% CI] \\
\midrule
Mutation score & 0.746 & 0.682 & 0.651 & 0.537 & 0.156 [0.106, 0.207] \\
Line coverage & 0.894 & 0.869 & 0.847 & 0.742 & 0.129 [0.077, 0.180] \\
Coverage at pass & 0.894 & 0.869 & 0.847 & 0.742 & 0.129 [0.077, 0.180] \\
Mutants killed per test & 0.38 & 0.53 & 1.96 & 21.28 & $-$20.56 [$-$26.87, $-$14.25] \\
Assertion density & 0.257 & 0.311 & 0.314 & 0.145 & 0.149 [0.117, 0.182] \\
Test redundancy & -- & -- & -- & -- & -- \\
Invalid-test rate & 0.0001 & 0.0000 & 0.0000 & 0.0000 & 0.0000 [$-$0.0000, 0.0001] \\
Stub sensitivity & 0.855 & 0.852 & 0.922 & 0.899 & $-$0.004 [$-$0.091, 0.082] \\
Vacuous test rate & 0.145 & 0.148 & 0.078 & 0.101 & 0.004 [$-$0.082, 0.091] \\
Tests written & 219 & 123 & 31 & 49 & 74 [42, 107] \\
Test lines of code & 2459 & 1406 & 494 & 1443 & 10 [$-$594, 613] \\
Test lines added & 2458 & 1406 & 494 & -- & -- \\
Assertions written & 623 & 434 & 157 & 140 & 264 [195, 334] \\
Mutants killed per added test & 0.135 & 0.176 & 0.591 & -- & -- \\
Mutants killed per added test line & 0.0125 & 0.0159 & 0.0387 & -- & -- \\
Generation seconds & 1384 & 551 & 110 & -- & -- \\
Tool calls & 68 & 61 & 16 & -- & -- \\
Output tokens & 112268 & 43194 & 11763 & -- & -- \\
Input tokens & 6437114 & 1524567 & 607467 & -- & -- \\
Mutation score per minute & 0.058 & 0.095 & 0.451 & -- & -- \\
Mutation score per tool call & 0.0204 & 0.0153 & 0.0571 & -- & -- \\
Suite runtime seconds & 4.90 & 2.02 & 0.61 & 7.98 & $-$5.47 [$-$13.15, 2.20] \\
Behavioral coverage & 0.887 & 0.872 & 0.796 & 0.599 & 0.253 [0.185, 0.321] \\
Edge-case coverage & 0.864 & 0.803 & 0.695 & 0.385 & 0.402 [0.343, 0.462] \\
Behaviors per test & 0.056 & 0.102 & 0.333 & 2.747 & $-$2.612 [$-$3.688, $-$1.536] \\
Edge cases per test & 0.140 & 0.231 & 0.682 & 4.091 & $-$3.781 [$-$5.270, $-$2.292] \\
\bottomrule
\end{tabular}
\end{table}

\newpage

\begin{table}[ht]
\centering
\small
\caption{Rust projects: mean of each test-suite author on every metric, and the paired gap between the agents and the human baseline. ``Agent mean'' averages Claude, Cursor, and Codex on the same project.}
\label{tab:metrics-rust}
\begin{tabular}{lrrrrl}
\toprule
 & \multicolumn{4}{c}{Mean by author} & \\
\cmidrule(lr){2-5}
Metric & Claude & Cursor & Codex & Human & Agent mean $-$ Human [95\% CI] \\
\midrule
Mutation score & 0.816 & 0.804 & 0.730 & 0.632 & 0.151 [0.114, 0.188] \\
Line coverage & 0.904 & 0.906 & 0.841 & 0.701 & 0.182 [0.141, 0.224] \\
Coverage at pass & 0.904 & 0.906 & 0.841 & 0.701 & 0.182 [0.141, 0.224] \\
Mutants killed per test & 0.39 & 0.51 & 1.71 & 2.51 & $-$1.64 [$-$2.36, $-$0.92] \\
Assertion density & 0.200 & 0.239 & 0.248 & 0.168 & 0.061 [0.036, 0.087] \\
Test redundancy & -- & -- & -- & -- & -- \\
Invalid-test rate & 0.0000 & 0.0050 & 0.0000 & 0.0000 & 0.0017 [$-$0.0007, 0.0040] \\
Stub sensitivity & 0.848 & 0.855 & 0.886 & 0.868 & 0.001 [$-$0.031, 0.033] \\
Vacuous test rate & 0.152 & 0.145 & 0.114 & 0.132 & $-$0.001 [$-$0.033, 0.031] \\
Tests written & 165 & 118 & 33 & 60 & 46 [17, 74] \\
Test lines of code & 2110 & 1678 & 515 & 859 & 575 [415, 735] \\
Test lines added & 2110 & 1678 & 515 & -- & -- \\
Assertions written & 416 & 391 & 132 & 139 & 174 [123, 225] \\
Mutants killed per added test & 0.088 & 0.116 & 0.301 & -- & -- \\
Mutants killed per added test line & 0.0064 & 0.0076 & 0.0181 & -- & -- \\
Generation seconds & 1039 & 618 & 119 & -- & -- \\
Tool calls & 49 & 68 & 18 & -- & -- \\
Output tokens & 84648 & 51087 & 12658 & -- & -- \\
Input tokens & 4577985 & 1903550 & 762791 & -- & -- \\
Mutation score per minute & 0.073 & 0.095 & 0.448 & -- & -- \\
Mutation score per tool call & 0.0281 & 0.0164 & 0.0499 & -- & -- \\
Suite runtime seconds & 0.97 & 0.75 & 0.56 & 2.91 & $-$2.15 [$-$4.30, 0.01] \\
Behavioral coverage & 0.884 & 0.883 & 0.819 & 0.570 & 0.292 [0.213, 0.371] \\
Edge-case coverage & 0.856 & 0.832 & 0.692 & 0.476 & 0.317 [0.259, 0.375] \\
Behaviors per test & 0.051 & 0.078 & 0.237 & 0.237 & $-$0.115 [$-$0.179, $-$0.050] \\
Edge cases per test & 0.128 & 0.182 & 0.508 & 0.457 & $-$0.184 [$-$0.316, $-$0.052] \\
\bottomrule
\end{tabular}
\end{table}

\newpage

\begin{table}[ht]
\centering
\small
\caption{Small projects: mean of each test-suite author on every metric, and the paired gap between the agents and the human baseline. ``Agent mean'' averages Claude, Cursor, and Codex on the same project.}
\label{tab:metrics-small}
\begin{tabular}{lrrrrl}
\toprule
 & \multicolumn{4}{c}{Mean by author} & \\
\cmidrule(lr){2-5}
Metric & Claude & Cursor & Codex & Human & Agent mean $-$ Human [95\% CI] \\
\midrule
Mutation score & 0.863 & 0.852 & 0.837 & 0.674 & 0.176 [0.134, 0.218] \\
Line coverage & 0.930 & 0.927 & 0.935 & 0.808 & 0.123 [0.088, 0.158] \\
Coverage at pass & 0.930 & 0.927 & 0.935 & 0.808 & 0.123 [0.088, 0.158] \\
Mutants killed per test & 0.36 & 0.44 & 1.66 & 11.28 & $-$10.56 [$-$14.96, $-$6.17] \\
Assertion density & 0.234 & 0.277 & 0.253 & 0.187 & 0.068 [0.038, 0.097] \\
Test redundancy & 0.254 & 0.266 & 0.218 & 0.303 & $-$0.059 [$-$0.130, 0.012] \\
Invalid-test rate & 0.0000 & 0.0024 & 0.0000 & 0.0000 & 0.0008 [$-$0.0008, 0.0023] \\
Stub sensitivity & 0.866 & 0.896 & 0.917 & 0.889 & 0.015 [$-$0.037, 0.067] \\
Vacuous test rate & 0.134 & 0.104 & 0.083 & 0.111 & $-$0.015 [$-$0.067, 0.037] \\
Tests written & 152 & 124 & 32 & 37 & 65 [42, 89] \\
Test lines of code & 1091 & 933 & 317 & 299 & 482 [388, 575] \\
Test lines added & 1091 & 933 & 317 & -- & -- \\
Assertions written & 255 & 255 & 82 & 51 & 147 [130, 164] \\
Mutants killed per added test & 0.086 & 0.108 & 0.356 & -- & -- \\
Mutants killed per added test line & 0.0111 & 0.0128 & 0.0334 & -- & -- \\
Generation seconds & 569 & 444 & 87 & -- & -- \\
Tool calls & 29 & 42 & 12 & -- & -- \\
Output tokens & 44099 & 35150 & 8668 & -- & -- \\
Input tokens & 1987904 & 796060 & 295296 & -- & -- \\
Mutation score per minute & 0.119 & 0.137 & 0.676 & -- & -- \\
Mutation score per tool call & 0.0380 & 0.0241 & 0.0793 & -- & -- \\
Suite runtime seconds & 1.53 & 1.25 & 0.73 & 1.64 & $-$0.47 [$-$1.37, 0.44] \\
Behavioral coverage & 0.935 & 0.928 & 0.893 & 0.709 & 0.209 [0.138, 0.281] \\
Edge-case coverage & 0.935 & 0.917 & 0.861 & 0.490 & 0.414 [0.350, 0.479] \\
Behaviors per test & 0.034 & 0.043 & 0.144 & 1.375 & $-$1.312 [$-$2.066, $-$0.558] \\
Edge cases per test & 0.099 & 0.124 & 0.408 & 1.659 & $-$1.458 [$-$2.190, $-$0.727] \\
\bottomrule
\end{tabular}
\end{table}

\newpage

\begin{table}[ht]
\centering
\small
\caption{Medium projects: mean of each test-suite author on every metric, and the paired gap between the agents and the human baseline. ``Agent mean'' averages Claude, Cursor, and Codex on the same project.}
\label{tab:metrics-medium}
\begin{tabular}{lrrrrl}
\toprule
 & \multicolumn{4}{c}{Mean by author} & \\
\cmidrule(lr){2-5}
Metric & Claude & Cursor & Codex & Human & Agent mean $-$ Human [95\% CI] \\
\midrule
Mutation score & 0.817 & 0.786 & 0.731 & 0.616 & 0.162 [0.116, 0.208] \\
Line coverage & 0.937 & 0.923 & 0.890 & 0.738 & 0.179 [0.132, 0.225] \\
Coverage at pass & 0.937 & 0.923 & 0.890 & 0.738 & 0.179 [0.132, 0.225] \\
Mutants killed per test & 0.36 & 0.54 & 1.72 & 8.83 & $-$7.95 [$-$12.22, $-$3.68] \\
Assertion density & 0.243 & 0.303 & 0.284 & 0.165 & 0.112 [0.083, 0.141] \\
Test redundancy & 0.126 & 0.144 & 0.126 & 0.240 & $-$0.108 [$-$0.149, $-$0.068] \\
Invalid-test rate & 0.0001 & 0.0021 & 0.0000 & 0.0000 & 0.0007 [$-$0.0007, 0.0021] \\
Stub sensitivity & 0.867 & 0.851 & 0.921 & 0.916 & $-$0.031 [$-$0.063, 0.001] \\
Vacuous test rate & 0.133 & 0.149 & 0.079 & 0.084 & 0.031 [$-$0.001, 0.063] \\
Tests written & 309 & 167 & 54 & 162 & 15 [$-$100, 129] \\
Test lines of code & 2242 & 1638 & 594 & 1312 & 180 [$-$228, 587] \\
Test lines added & 2242 & 1638 & 594 & -- & -- \\
Assertions written & 535 & 485 & 168 & 202 & 194 [130, 257] \\
Mutants killed per added test & 0.110 & 0.147 & 0.445 & -- & -- \\
Mutants killed per added test line & 0.0117 & 0.0141 & 0.0315 & -- & -- \\
Generation seconds & 1149 & 665 & 126 & -- & -- \\
Tool calls & 61 & 68 & 18 & -- & -- \\
Output tokens & 93507 & 54062 & 13900 & -- & -- \\
Input tokens & 5992566 & 1893812 & 715974 & -- & -- \\
Mutation score per minute & 0.051 & 0.077 & 0.372 & -- & -- \\
Mutation score per tool call & 0.0175 & 0.0130 & 0.0467 & -- & -- \\
Suite runtime seconds & 3.12 & 1.78 & 0.74 & 5.01 & $-$3.13 [$-$5.66, $-$0.61] \\
Behavioral coverage & 0.898 & 0.888 & 0.832 & 0.536 & 0.336 [0.274, 0.398] \\
Edge-case coverage & 0.886 & 0.850 & 0.711 & 0.455 & 0.361 [0.309, 0.414] \\
Behaviors per test & 0.050 & 0.088 & 0.270 & 0.967 & $-$0.831 [$-$1.364, $-$0.298] \\
Edge cases per test & 0.123 & 0.207 & 0.553 & 1.757 & $-$1.462 [$-$2.511, $-$0.414] \\
\bottomrule
\end{tabular}
\end{table}

\newpage

\begin{table}[ht]
\centering
\small
\caption{Large projects: mean of each test-suite author on every metric, and the paired gap between the agents and the human baseline. ``Agent mean'' averages Claude, Cursor, and Codex on the same project.}
\label{tab:metrics-large}
\begin{tabular}{lrrrrl}
\toprule
 & \multicolumn{4}{c}{Mean by author} & \\
\cmidrule(lr){2-5}
Metric & Claude & Cursor & Codex & Human & Agent mean $-$ Human [95\% CI] \\
\midrule
Mutation score & 0.752 & 0.647 & 0.546 & 0.566 & 0.083 [0.029, 0.137] \\
Line coverage & 0.905 & 0.875 & 0.755 & 0.730 & 0.115 [0.057, 0.172] \\
Coverage at pass & 0.905 & 0.875 & 0.755 & 0.730 & 0.115 [0.057, 0.172] \\
Mutants killed per test & 0.24 & 0.44 & 1.29 & 3.43 & $-$2.78 [$-$5.31, $-$0.24] \\
Assertion density & 0.243 & 0.286 & 0.317 & 0.154 & 0.128 [0.100, 0.157] \\
Test redundancy & 0.071 & 0.087 & 0.082 & 0.206 & $-$0.125 [$-$0.192, $-$0.059] \\
Invalid-test rate & 0.0000 & 0.0000 & 0.0000 & 0.0000 & 0.0000 [0.0000, 0.0000] \\
Stub sensitivity & 0.862 & 0.863 & 0.921 & 0.864 & 0.049 [$-$0.027, 0.126] \\
Vacuous test rate & 0.138 & 0.137 & 0.079 & 0.136 & $-$0.049 [$-$0.126, 0.027] \\
Tests written & 503 & 210 & 58 & 752 & $-$495 [$-$1504, 514] \\
Test lines of code & 4110 & 2442 & 777 & 3522 & $-$1079 [$-$2133, $-$26] \\
Test lines added & 4108 & 2442 & 777 & -- & -- \\
Assertions written & 986 & 679 & 238 & 488 & 147 [18, 276] \\
Mutants killed per added test & 0.074 & 0.121 & 0.290 & -- & -- \\
Mutants killed per added test line & 0.0062 & 0.0077 & 0.0179 & -- & -- \\
Generation seconds & 2258 & 780 & 164 & -- & -- \\
Tool calls & 136 & 105 & 27 & -- & -- \\
Output tokens & 184685 & 66117 & 17928 & -- & -- \\
Input tokens & 14026809 & 3572865 & 1537231 & -- & -- \\
Mutation score per minute & 0.024 & 0.053 & 0.212 & -- & -- \\
Mutation score per tool call & 0.0075 & 0.0070 & 0.0229 & -- & -- \\
Suite runtime seconds & 6.00 & 3.23 & 1.01 & 17.48 & $-$14.06 [$-$26.59, $-$1.54] \\
Behavioral coverage & 0.841 & 0.834 & 0.720 & 0.539 & 0.260 [0.194, 0.326] \\
Edge-case coverage & 0.815 & 0.723 & 0.533 & 0.431 & 0.259 [0.199, 0.320] \\
Behaviors per test & 0.062 & 0.140 & 0.418 & 0.745 & $-$0.538 [$-$1.244, 0.168] \\
Edge cases per test & 0.151 & 0.297 & 0.761 & 1.316 & $-$0.913 [$-$1.919, 0.093] \\
\bottomrule
\end{tabular}
\end{table}

\newpage
\section{Prompts}
\label{prompts}
\begin{lstlisting}[
  caption={Test suite generation prompt (C++).},
  label={lst:test-generation-prompt}
]
        Write a unit test suite for this project.

        The production code is under `src/`. Put your tests under
        `tests/`. Do not change anything under `src/`: the
        task is to test the code as it is, and an edit there invalidates the run.

        `tests/` holds one file, `tests/CMakeLists.txt`, and it belongs to
        the harness rather than to you: the build system names it, so deleting it or
        writing tests into it stops the project building. Leave it exactly as it is and
        add your tests in new files beside it.

        Every `.cpp` file you add under `tests/` is compiled into its own test
        executable, and every `TEST(...)` in it is registered with CTest for you,
        including in a subdirectory you create. You do not need to add anything to
        `tests/CMakeLists.txt`. GoogleTest and GoogleMock are available and
        linked for you, including their `main`, so write `TEST(...)` and do not define
        `main` yourself.

        Dependencies are already installed and the project already builds. There is no
        network: nothing can be installed, fetched, or looked up, so write the suite
        from the code in front of you. Run your test suite with:

            cmake --build build && HTTP_PROXY= HTTPS_PROXY= http_proxy= https_proxy= \
                ctest --test-dir build --no-tests=error

        Write the suite a maintainer would accept. Cover the public behavior of each
        module, including error paths and edge cases, and make every assertion
        one that would fail if the behavior it describes changed. Do not write tests
        that assert nothing, or that only assert a call did not raise.

        Stop when you have finished writing your tests and the whole suite passes.
\end{lstlisting}

\begin{lstlisting}[
  caption={Transpilation task prompt.},
  label={lst:transpilation-prompt}
]
        This directory contains a rust project whose source was translated from
        the project in `source_proj/`. Translate each test from `source_proj/`
        into rust.

        The production code is under `src/`. Put your tests under
        `tests/`. Do not change anything under `src/` or
        `source_proj/`: the task is to test the code as it is, and an edit there
        invalidates the run.

        `tests/` is empty.

        Cargo builds an integration test from `tests/<name>.rs` and from
        `tests/<dir>/main.rs`, and from nowhere else. A file in a
        subdirectory is a module, not a target, so a test written to
        `tests/unit/parser.rs` is never compiled and never runs unless a
        `main.rs` beside it declares it.

        Dependencies are already installed and the project already builds. There is no
        network: nothing can be installed, fetched, or looked up, so write the suite
        from the code in front of you. Run your test suite with:

            HTTP_PROXY= HTTPS_PROXY= http_proxy= https_proxy= \
                cargo nextest run --workspace

        Write the suite a maintainer would accept. Reproduce the behavior of each test
        in `source_proj/` as faithfully as you can. You may skip any tests whose
        translation is impossible because the translated rust implementation
        differs from the original behavior. Do not write tests that assert
        nothing, or that only assert a call did not raise.

        Stop when you have translated each test into rust and the whole suite
        passes.
\end{lstlisting}

\newpage 

\begin{lstlisting}[
  caption={Failure-aware instructions used in the treatment prompt.},
  label={lst:failure-aware-prompt},
  basicstyle=\ttfamily\small,
  breaklines=true,
  columns=fullflexible
]
      Use this workflow:

      1. Inventory all production source files and public entry points before
         choosing what to test. Do not limit inspection to a few
         representative modules.

      2. For each selected behavior, construct inputs that actually reach the
         intended path. If setup is difficult, inspect the implementation and
         dependencies and repair the setup rather than silently replacing the
         behavior with an easier one.

      3. Derive expected results from an explicit contract, invariant, or
         independent calculation. Do not change an expected value solely to
         match observed output. If implementation behavior conflicts with an
         explicit contract, do not encode that behavior as correct merely to
         make the suite pass.

      4. Once the suite is green, measure source coverage with:

             <language-specific coverage command>

         Inspect the file-level report and missing lines or regions, then add
         meaningful tests for important uncovered behavior. Rerun both the
         suite and coverage after revisions. Passing tests alone are not
         evidence that the suite is complete.

      Stop when the whole suite passes and you have used the coverage report
      to check for important behavior that remains untested.
\end{lstlisting}

\section{Trajectory evidence}
\label{app:evidence}
\renewcommand{\thelstlisting}{C.\arabic{lstlisting}}
The examples below illustrate the findings using runs from the 675 GPT trajectories analyzed above. Each example names the project and identifies which of its three runs is shown. We reproduce the relevant agent statements, commands, execution output, source code, and generated tests so that readers can follow the evidence for each finding. Event identifiers locate the excerpts within the recorded run.

The excerpts preserve the recorded content. Long lines are wrapped to fit the page, and test-code excerpts omit the leading \texttt{+} symbols that marked added lines in the recorded diff.

\subsection{Unread implementation: minisketch (repetition C)}
The recorded source reads cover the public interface, dispatch, shared sketch algorithms, field configuration, and utility code. The complete command inventory contains no read of the eight generic field-backend implementation bodies. Their relevance is established by the recorded build configuration and the final tests, which exercise all eight storage widths, rather than inferred from a low metric. This supports a finding of omitted source inspection; it does not establish that these files were wholly untested.

\begin{lstlisting}[caption={Minisketch's first implementation-reading command.},label={lst:scope-commands}]
/bin/sh -lc "sed -n '1,260p' src/include/minisketch.h && sed -n '1,320p' src/minisketch.cpp && sed -n '1,280p' src/sketch.h && sed -n '1,360p' src/sketch_impl.h"
\end{lstlisting}

\begin{table}[h]
\small
\centering
\begin{tabular}{@{}p{32mm}p{103mm}@{}}
\toprule
Event & Other recorded source reads (analyst summary)\\
\midrule
\texttt{item\_3} & Further public-header and dispatcher reads; \texttt{false\_positives.h}; project, source, and test CMake files.\\
\texttt{item\_6} & Remaining shared sketch code; \texttt{fielddefines.h}, \texttt{util.h}, and \texttt{int\_utils.h}.\\
\texttt{item\_7} & Public header read from \texttt{HEAD}.\\
\bottomrule
\end{tabular}
\caption{Additional source inspection in the minisketch trajectory, summarized from the recorded commands.}
\end{table}

\newpage 

\begin{lstlisting}[caption={Captured build wiring includes all eight generic field-backend files.},label={lst:scope-build}]
add_library(minisketch_field_sources INTERFACE)
target_sources(minisketch_field_sources
  INTERFACE
    fields/generic_1byte.cpp
    fields/generic_2bytes.cpp
    fields/generic_3bytes.cpp
    fields/generic_4bytes.cpp
    fields/generic_5bytes.cpp
    fields/generic_6bytes.cpp
    fields/generic_7bytes.cpp
    fields/generic_8bytes.cpp
)
\end{lstlisting}

\noindent\textbf{Minisketch: final-test evidence (continued)}
\begin{lstlisting}[caption={The final minisketch tests exercise the corresponding storage widths.},label={lst:scope-test}]
TEST(Decoding, WorksAcrossEveryFieldStorageWidth)
{
    const uint32_t representative_bits[] = {8, 16, 24, 32, 40, 48, 56, 64};
    for (uint32_t bits : representative_bits) {
        for (uint32_t implementation = 0;
             implementation <= minisketch_implementation_max(); ++implementation) {
            if (!minisketch_implementation_supported(bits, implementation)) continue;
            auto sketch = MakeSketch(bits, implementation, 3);
            ASSERT_NE(nullptr, sketch) << bits << "/" << implementation;
            const uint64_t high = bits == 64 ? (uint64_t{1} << 63)
                                             : (uint64_t{1} << (bits - 1));
            minisketch_add_uint64(sketch.get(), 1);
            minisketch_add_uint64(sketch.get(), high);
            minisketch_add_uint64(sketch.get(), high + 7);
            EXPECT_EQ(std::vector<uint64_t>({1, high, high + 7}),
                      Decode(sketch.get(), 3))
                << bits << "/" << implementation;
        }
    }
}
\end{lstlisting}

The source also links \texttt{minisketch\_field\_sources} into the \texttt{minisketch} library. Conditional CLMUL backends were disabled in this build; the finding concerns the active generic backends and does not depend on CLMUL. We checked the complete inventory of recorded reads to establish the omission; Listing~\ref{lst:scope-commands} shows one selected command.

\subsection{Retained area-formula error: geometer (repetition B)}
The agent explicitly identifies the circle-area formula as erroneous. The final test nevertheless requires that formula's result. For radius 2, the mathematical area is $4\pi$, while the retained assertion requires $8\pi$.

\begin{lstlisting}[caption={Recorded recognition of the circle-area error.},label={lst:area-recognition}]
There's this known issue with the calculation of a circle's area — it returns 2πr² instead of the expected πr². I should confirm this in the public behavior documentation and ensure it passes quality checks.
\end{lstlisting}

\newpage 

\begin{lstlisting}[caption={The final assertion retains the recognized erroneous result.},label={lst:area-test}]
def test_circle_lie_coordinates_intersection_angle_area_and_validation():
    unit = CircularCurve()
    np.testing.assert_allclose(unit.lie_coordinates, [0, 1, 0, 0])
    assert unit.intersection_angle(unit) == pytest.approx(0)
    assert CircularCurve(radius=2).area == pytest.approx(8 * np.pi)
\end{lstlisting}

The final run reports 104 passing tests (\texttt{item\_51}), but the area assertion does not establish the intended formula. The task required a passing suite and prohibited edits to production source code. Those constraints provide context for the choice without changing what the retained assertion requires.

\subsection{Weakened missing-input validation: pyflwdir (repetition C)}
The initial test expects a \texttt{ValueError} for missing inputs, and the captured implementation contains the corresponding guard. Execution instead fails later during arithmetic on arrays containing \texttt{None}. The agent changes the final test to require that incidental \texttt{TypeError}.

\begin{lstlisting}[caption={The original assertion requires the missing-input validation error.},label={lst:validation-before}]
        with pytest.raises(ValueError, match='"zs" and "rivdst"'):
>           flw.river_depth(q, width)
\end{lstlisting}

\begin{lstlisting}[caption={The captured source contains the intended validation guard.},label={lst:validation-guard}]
        if rivslp is None:
            if zs is None or rivdst is None:
                raise ValueError('"zs" and "rivdst" are required if "rivslp" is None.')
            dz = zs - self.downstream(zs)
\end{lstlisting}

\begin{lstlisting}[caption={The observed exception is an incidental subtraction error.},label={lst:validation-error}]
E           TypeError: unsupported operand type(s) for -: 'NoneType' and 'NoneType'

src/pyflwdir/flwdir.py:780: TypeError
\end{lstlisting}

\begin{lstlisting}[caption={The final test accepts the incidental error instead of requiring the validation guard.},label={lst:validation-after}]
def test_river_depth_manning_and_argument_errors(flw):
    q = np.full(flw.shape, 100.0)
    width = np.full(flw.shape, 10.0)
    slope = np.full(flw.shape, 0.001)
    depth = flw.river_depth(q, width, rivslp=slope, manning=0.03)
    expected = ((0.03 * 100) / (np.sqrt(0.001) * 10)) ** (3 / 5)
    np.testing.assert_allclose(depth, expected)
    with pytest.raises(ValueError, match="Method unknown"):
        flw.river_depth(q, width, rivslp=slope, method="other")
    with pytest.raises(TypeError, match="unsupported operand"):
        flw.river_depth(q, width)
\end{lstlisting}

In this trajectory, accepting the output from malformed input also weakens the intended validation oracle. The finding rests on the change in the assertion and does not require an unrecorded admission of a bug. The source guard supports the intended exception; the docstring describes the required inputs but does not independently specify an exception class. The final suite reports 59 passing tests (\texttt{item\_47}).

\clearpage
\subsection{Execution without coverage measurement: geometer (repetition C)}
This run is a different repetition from the oracle example above. The record contains four completed suite executions using the same command. The number of failures falls until the suite passes. Review of the complete trajectory identifies no coverage-measurement attempt or inherited instrumentation.

\begin{lstlisting}[caption={The command used for each of the four geometer suite executions.},label={lst:feedback-command}]
/bin/sh -lc "HTTP_PROXY= HTTPS_PROXY= http_proxy= https_proxy= PYTHONPATH=/work/src PYTHONDONTWRITEBYTECODE=1 python -m pytest tests -p no:cacheprovider -o addopts= -o log_file= -o 'python_files=test*.py *_test.py *_tests.py' --continue-on-collection-errors"
\end{lstlisting}

\begin{table}[h]
\centering
\small
\begin{tabular}{@{}lrrr@{}}
\toprule
Event & Failed & Passed & Exit code\\
\midrule
\texttt{item\_22} & 16 & 77 & 1\\
\texttt{item\_28} & 3 & 90 & 1\\
\texttt{item\_30} & 1 & 92 & 1\\
\texttt{item\_34} & 0 & 93 & 0\\
\bottomrule
\end{tabular}
\caption{Recorded test-execution outcomes for geometer, repetition C.}
\label{tab:feedback-runs}
\end{table}

\begin{lstlisting}[caption={The final recorded pytest summary.},label={lst:feedback-result}]
============================== 93 passed in 0.39s ==============================
\end{lstlisting}

The initial file inventory lists coverage executables, but listing an executable is not a measurement attempt. To assess whether coverage was measured, we reviewed all recorded commands, configuration, and instrumentation; the selected pytest commands alone would not establish its absence. This example documents the feedback process without establishing weak final coverage or a causal effect on test quality.

\begin{table}[h]
\centering
\small
\begin{tabular}{@{}lrr@{}}
\toprule
Recorded coverage signal & Trajectories & Of all 675\\
\midrule
No measurement attempt & 376 & 55.7\%\\
Measurement attempted & 211 & 31.3\%\\
Successful measurement documented & 99 & 14.7\%\\
\bottomrule
\end{tabular}
\caption{Selected coverage observations, including inherited instrumentation. Rows are not an exhaustive partition; successful measurement is a subset of attempted measurement. No weak-outcome classification is used.}
\label{tab:coverage-observations}
\end{table}

\subsection{Stated familiarity}
In Listings~\ref{lst:familiarity-cleo}--\ref{lst:familiarity-more-itertools}, the agent describes familiarity with the libraries while planning tests. These statements document claimed familiarity; they do not independently establish prior knowledge. We do not assess a relationship between these statements and test quality here.

In Cleo C, betfairlightweight B, pycasbin B, and PyJWT B and C, the statements precede the first recorded read of implementation or test contents. The preceding commands inspect file inventories, repository metadata, or project configuration. The betfairlightweight, pycasbin, and PyJWT B excerpts specifically claim knowledge of upstream tests, whereas Cleo C and PyJWT C describe familiarity with the project.

\begin{lstlisting}[caption={Stated familiarity with Cleo, repetition C.},label={lst:familiarity-cleo}]
It seems like Cleo is an open-source project I'm familiar with.
\end{lstlisting}

\begin{lstlisting}[caption={Stated familiarity with upstream tests in betfairlightweight, repetition B, before implementation or test-content inspection.},label={lst:familiarity-betfairlightweight}]
I'm familiar with some tests from the betfairlightweight GitHub repository, but there's no network access.
\end{lstlisting}

\begin{lstlisting}[caption={Stated knowledge of pycasbin tests, repetition B, before implementation or test-content inspection.},label={lst:familiarity-pycasbin}]
I might also leverage what I know about pycasbin tests.
\end{lstlisting}

\begin{lstlisting}[caption={Stated knowledge of upstream tests in PyJWT, repetition B, before implementation or test-content inspection.},label={lst:familiarity-pyjwt-b}]
I could leverage what I know from the upstream tests.
\end{lstlisting}

\begin{lstlisting}[caption={Stated existing familiarity with PyJWT, repetition C, before implementation or test-content inspection.},label={lst:familiarity-pyjwt-c}]
It looks like I could use the PyJWT version 2.13 since I have some existing familiarity with it.
\end{lstlisting}

The more-itertools C statement below follows source inspection. It documents a claim of familiarity, but the trace does not establish whether that familiarity predates the run.

\begin{lstlisting}[caption={Stated familiarity and plans to draw on that knowledge in more-itertools, repetition C.},label={lst:familiarity-more-itertools}]
I'll inspect implementations in manageable chunks and leverage the library I'm familiar with. Writing tests based on this knowledge will be critical.
\end{lstlisting}

\newpage
\section{Formulas Used}
\label{app:metrics}

\begin{table}[h]
\centering
\caption{Notation.}
\label{tab:symbols}
\small
\begin{tabular}{@{}ll@{}}
\toprule
Symbol & Meaning \\
\midrule
$U$
  & Production lines we measure. \\
$\mathrm{Cover}(P)$
  & Those lines that passing tests execute. \\
$K$, $V$
  & Killed mutants and viable mutants. \\
$R$
  & Viable mutants the suite actually reaches. \\
$n$, $n_{\mathrm{loc}}$
  & Tests written, and test lines written. \\
$T_t$
  & Lines test $t$ covers, after removing shared setup. \\
$A$, $\mathit{Sens}$
  & Tests that pass on the original code; and the subset of those that then fail, hang, or crash against the stub. \\
$\mathcal{B}$, $\mathcal{E}$
  & Catalog behaviors and edge cases. \\
\bottomrule
\end{tabular}
\end{table}

\begin{table}[h]
\centering
\caption{Reported metrics. $\uparrow$ higher is better; $\downarrow$ lower is better.}
\label{tab:all-metrics}
\small
\begin{tabular}{@{}lcl@{}}
\toprule
Metric & & Formula \\
\midrule
Line coverage
  & $\uparrow$
  & $\lvert\mathrm{Cover}(P)\rvert/\lvert U\rvert$ \\
Mutation score
  & $\uparrow$
  & $\lvert K\rvert/\lvert V\rvert$ \\
Kill rate (reached)
  & $\uparrow$
  & $\lvert K\rvert/\lvert R\rvert$ \\
Mutants not reached
  & $\downarrow$
  & $1-\lvert R\rvert/\lvert V\rvert$ \\
Tests written
  & --
  & $n$ \\
Test lines
  & --
  & $n_{\mathrm{loc}}$ \\
Assertion density
  & $\uparrow$
  & assertions $/n_{\mathrm{loc}}$ \\
Stub sensitivity
  & $\uparrow$
  & $\lvert\mathit{Sens}\rvert/\lvert A\rvert$ \\
Redundancy
  & $\downarrow$
  & mean overlap of $T_t$ across pairs of tests \\
Vacuous-test rate
  & $\downarrow$
  & $\lvert A\setminus\mathit{Sens}\rvert/\lvert A\rvert$ \\
Invalid-test rate
  & $\downarrow$
  & tests that do not pass $/$ all tests \\
Kills per added test
  & $\uparrow$
  & new kills $/n$ \\
Kills per added line
  & $\uparrow$
  & new kills $/n_{\mathrm{loc}}$ \\
Faults caught per test
  & $\uparrow$
  & $\lvert K\rvert/n$ \\
Behavioral coverage
  & $\uparrow$
  & behaviors hit $/\lvert\mathcal{B}\rvert$ \\
Edge case coverage
  & $\uparrow$
  & edge cases hit $/\lvert\mathcal{E}\rvert$ \\
Behaviors per test
  & $\uparrow$
  & behaviors hit $/n$ \\
Edge cases per test
  & $\uparrow$
  & edge cases hit $/n$ \\
Generation time
  & $\downarrow$
  & wall-clock seconds \\
Tool calls
  & $\downarrow$
  & reported tool-call count \\
Output tokens
  & $\downarrow$
  & reported completion tokens \\
Cost (USD)
  & $\downarrow$
  & reported dollar cost \\
Mutation per minute
  & $\uparrow$
  & mutation score $/$ generation minutes \\
Mutation per tool call
  & $\uparrow$
  & mutation score $/$ tool calls \\
\bottomrule
\end{tabular}
\end{table}

\section{Failure aware prompt experiment}
\label{sec:failure-aware-prompt-experiment}
\begin{table}[H]
\centering
\caption{Test-quality metrics on the matched 41-project cohort. The standard-prompt value is the mean of three runs; the failure-aware prompt uses one run.}
\label{tab:failure-aware-prompt-comparison}
\resizebox{\textwidth}{!}{%
\begin{tabular}{lrrr}
\toprule
\textbf{Metric}
& \textbf{Standard prompt}
& \textbf{Failure-aware prompt}
& \textbf{Difference} \\
\midrule
Line coverage 
& 92.21\%
& \textbf{96.43\%}
& $+4.22$ pp \\

Mutation score 
& 69.98\%
& \textbf{73.61\%}
& $+3.63$ pp \\

Redundancy 
& \textbf{18.01\%}
& 20.80\%
& $+2.79$ pp \\

Vacuous-test rate 
& 3.86\%
& \textbf{3.62\%}
& $-0.24$ pp \\
\bottomrule
\end{tabular}%
}
\end{table}

\section{Compute Used}
All experiments were run on 7 AWS EC2 \texttt{c8a.16xlarge} CPU instances
for approximately 5 days, corresponding to roughly 840 instance-hours of
compute. Each instance provided 64 AMD x86-64 CPU cores (64 vCPUs,
single-threaded with no SMT) and 128 GiB of memory, for a maximum concurrent
allocation of 448 CPU cores and 896 GiB of memory. No GPUs were used.

\newpage
\section{Data Provenance}
{\scriptsize\setlength{\tabcolsep}{4pt}
\begin{longtable}{@{}>{\raggedright\arraybackslash}p{2.3cm}>{\raggedright\arraybackslash}p{3.0cm}>{\raggedright\arraybackslash\ttfamily}p{1.4cm}>{\raggedright\arraybackslash}p{1.3cm}>{\raggedright\arraybackslash}p{4.8cm}@{}}
\label{tab:dataset-projects}\\
\toprule
Project & Repository & Commit & License & Citation \\
\midrule
\endfirsthead
\toprule
Project & Repository & Commit & License & Citation \\
\midrule
\endhead
\bottomrule
\endfoot
\midrule\multicolumn{5}{l}{\textit{C++ projects}} \\
AHRS & \href{https://github.com/wgrand/AHRS}{\nolinkurl{wgrand/AHRS}} & a9f989d07b\allowbreak{}f1acc75774\allowbreak{}f5e595c265\allowbreak{}2eed9385f9 & MIT & William Grand. AHRS. \url{https://github.com/wgrand/AHRS} (commit a9f989d07bf1acc75774f5e595c2652eed9385f9). \\
artificial-neural-network & \href{https://github.com/camchambers/artificial-neural-network}{\nolinkurl{camchambers/artificial-neural-network}} & f74dc79cda\allowbreak{}62ff7f0494\allowbreak{}88cb3cf00c\allowbreak{}25fdf63e0e & MIT & Cam. artificial-neural-network. \url{https://github.com/camchambers/artificial-neural-network} (commit f74dc79cda62ff7f049488cb3cf00c25fdf63e0e). \\
base64 & \href{https://github.com/tobiaslocker/base64}{\nolinkurl{tobiaslocker/base64}} & 8d96a2a737\allowbreak{}ac1396304b\allowbreak{}1de289beb3\allowbreak{}a5ea0cb752 & MIT & Tobias Locker. base64. \url{https://github.com/tobiaslocker/base64} (commit 8d96a2a737ac1396304b1de289beb3a5ea0cb752). \\
biscuit & \href{https://github.com/lioncash/biscuit}{\nolinkurl{lioncash/biscuit}} & 805c1756dd\allowbreak{}602e50bd4f\allowbreak{}57e772998a\allowbreak{}b3ad91df7f & MIT & Lioncash/Lioncache. biscuit. \url{https://github.com/lioncash/biscuit} (commit 805c1756dd602e50bd4f57e772998ab3ad91df7f). \\
Bit-Manipulation & \href{https://github.com/Zhuagenborn/Bit-Manipulation}{\nolinkurl{Zhuagenborn/Bit-Manipulation}} & 365848f2bc\allowbreak{}366dafc2e9\allowbreak{}d008bea693\allowbreak{}22bd3d186f & MIT & Zhenshuo Chen and Guowen Liu. Bit Manipulation. 2024. \url{https://github.com/Zhuagenborn/Bit-Manipulation} (commit 365848f2bc366dafc2e9d008bea69322bd3d186f). \\
blazecsv & \href{https://github.com/juanchristensen/blazecsv}{\nolinkurl{juanchristensen/blazecsv}} & 021eaa6841\allowbreak{}f202434884\allowbreak{}ab971e4a37\allowbreak{}6700ea98c5 & MIT & BlazeCSV Contributors. blazecsv. \url{https://github.com/juanchristensen/blazecsv} (commit 021eaa6841f202434884ab971e4a376700ea98c5). \\
\end{longtable}
}

\section{Statement of LLM Usage}
We used LLM-based tools for limited assistance during the preparation of this work. A coding agent was leveraged to assist with aspects of implementation after the experimental design and intended behavior had been specified by the authors. LLMs were also used as editorial aids to review drafts for clarity, concision, and phrasing. All methodological decisions, experimental design, analysis, and scientific conclusions were made by the authors, who reviewed and verified the resulting code and manuscript.

\end{document}